\documentclass[fleqn,twoside]{article}
\usepackage{espcrc2}
\usepackage{amssymb}

\usepackage{graphicx}
\usepackage[figuresright]{rotating}

\mathchardef\qsm=63
\mathchardef\pls=43
\mathchardef\mns=512
\mathchardef\plm=518
\mathchardef\eql=61
\mathchardef\smallleft=300
\mathchardef\smallright=301
\mathchardef\les=316
\mathchardef\gre=318
\mathchardef\leq=532
\mathchardef\grq=533

\newcommand {\IP}  {I\hspace{-0.2em}P}

\newcommand{\AmS}{{\protect\the\textfont2
  A\kern-.1667em\lower.5ex\hbox{M}\kern-.125emS}}

\title{
  QCD at high energy (experiments)
  \thanks{
    Talk presented at the $31^{\rm st}$ International Conference on
    High Energy Physics, Amsterdam, 24-31 July 2002.
  }
}

\author{K. Long
  \address[MCSD]{
    High Energy Physics Group, 
    Blackett Laboratory, 
    Department of Physics,
    Imperial College of Science Technology and Medicine,
    Prince Consort Road,
    London SW7 2BW,
    United Kingdom.
   }%
}

\begin{document}

\begin{abstract}
Recent measurements of QCD interactions involving large momentum
transfers are reviewed.
The status of measurements of the strong coupling constant is
summarised.
Recent developments in the measurement and interpretation of deep
inelastic scattering, proton-anti-proton collisions and two-photon
processes are discussed.
While QCD at next-to-leading order gives a qualitative description
of many processes, next-to-NLO calculations are now required to allow
quantitative information to be extracted from hadron-initiated
multijet data.
This is illustrated by a discussion of recent data on the
photoproduction of dijet events at HERA.
  \vspace{1pc}
\end{abstract}
\maketitle

\section{Introduction}
\label{Sect:Intro}

Experimentation over the last three decades has established
quantum chromodynamics (QCD) as the theory that describes the
interactions of quarks and gluons.
The phenomenology of QCD is rich, a consequence of the gluon
self-interactions.
These interactions cause the strong coupling, $\alpha_{\rm S}(\mu)$, to
fall with increasing momentum scale, $\mu$.
When $\mu$ is large, $\alpha_{\rm S}(\mu)$ is small and
perturbative QCD (pQCD) can be used to give an excellent description
of hadronic phenomena.
Conversely, as $\mu$ falls and $\alpha_{\rm S}(\mu)$ grows, higher-order
diagrams become increasingly important.
When $\mu$ is comparable to the masses of initial or final state
hadrons, the effects of confinement begin to become important.
In this kinematic regime, non-perturbative models must be employed to
obtain insight into the dynamics of the process under consideration.

A complete and quantitative understanding of QCD over the full
kinematic range is required before it can be claimed that the Standard
Model (SM) is understood.
For some processes, for example jet production in $e^+e^-$
annihilation, inclusive deep inelastic lepton-nucleon scattering and
inclusive jet production in $p \bar{p}$ and $ep$ scattering, the QCD
description is precise enough at next-to-leading order (NLO) for
quantitative information, such as the value of $\alpha_{\rm S}$, to
be extracted from the data.
The description of other processes is, at best, qualitative. 
Examples include multijet and heavy quark production
in $p \bar{p}$ and $ep$ scattering.
My purpose in this brief contribution is to review the data on hard
processes, i.e. those for which $\alpha_{\rm S}(\mu)$ is `small', and by
comparing the data to the pQCD predictions establish where 
quantitative information can be extracted and where progress
is required for the development of a full, quantitative understanding 
of QCD.

\section{Measurement of colour factors and $\boldmath \alpha_{\rm S}$}
\label{Sect:QCDParams}

\begin{figure}
  \begin{center}
    \includegraphics[width=0.7\columnwidth]{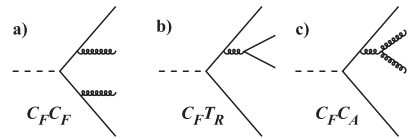}
  \end{center}
  \vspace{-0.8cm}
  \caption{
    A subset of the Feynman diagrams contributing at leading order to
    four-jet production in $e^+e^-$ annihilation.
    Each diagram is labelled with the combination of QCD colour
    factors upon which the matrix element depends.
  }
  \label{Fig:EE4JetFD}
\end{figure}
Electron-positron annihilation to hadrons at high energy is well
suited to the measurement of the parameters of QCD since the initial
state contains no hadrons.
Four-jet production in $e^+e^-$ annihilation is of particular
interest since at leading order (LO) the cross section is proportional
to $\alpha_{\rm S}^2$.
An illustrative subset of Feynman diagrams contributing at LO is
shown in figure \ref{Fig:EE4JetFD}.
Note that the gluon self interaction (figure \ref{Fig:EE4JetFD}c)
contributes at LO.
The `multiplicity' of the various colour configurations allowed by 
the theory is coded in the colour factors $C_F$, $C_A$, and $T_R$ as
indicated. 
The angular momentum of the initial states that may contribute are
determined by the vector nature of the electroweak interaction.
Hence, angular correlations amongst the final state jets may be used
to disentangle the various diagrams.

\begin{figure}
  \begin{center}
    \includegraphics[width=0.7\columnwidth]{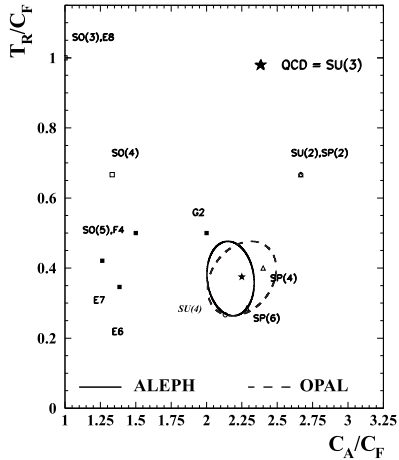}
  \end{center}
  \vspace{-0.8cm}
  \caption{
    68\% confidence level contours in the 
    $\left( C_A / C_F, T_R / C_F \right)$ plane obtained from
    measurements of angular correlations amongst the jets in 4-jet
    production at LEP.
    The ALEPH and OPAL results are shown as the solid and dashed lines
    respectively. 
  }
  \label{Fig:LEPCFSum}
\end{figure}

\begin{figure}
  \begin{center}
    \includegraphics[width=0.7\columnwidth]{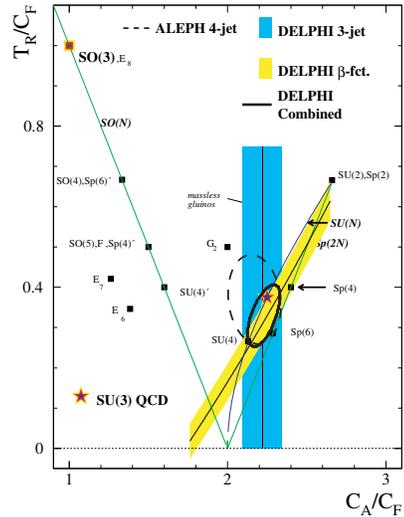}
  \end{center}
  \vspace{-0.8cm}
  \caption{
    Constraints in the $\left( C_A / C_F, T_R / C_F \right)$ plane
    obtained from analyses of the hadronic final states produced in
    $e^+ e^-$ annihilation.
    The vertical dark-shaded band shows the limit obtained
    from the ratio of multiplicity in gluon-induced to quark-induced
    jets. 
    The diagonal light-shaded band shows the constraint obtained
    from a measurement of the QCD $\beta$ function in a renormalisation 
    group invariant analysis.
    The solid ellipse shows the result obtained by combining the DELPHI
    measurements shown in the figure with the ALEPH measurement based
    on 4-jet angular correlations (dashed ellipse).
  }
  \label{Fig:LEPCFSum1}
\end{figure}

The LEP experiments ALEPH and OPAL have each made a
simultaneous fit to the four-jet rate and the distributions of the
angular correlations to determine simultaneously the colour factors
and $\alpha_{\rm S}(M_Z)$, where $M_Z$ is the mass of the $Z$ 
boson \cite{Ref:ALEPH4Jet,Ref:OPAL4Jet}.
The results are summarised in figure \ref{Fig:LEPCFSum}. 
The DELPHI collaboration has also presented such an analysis based on
a smaller data sample \cite{Ref:DELPHI4Jet}.
Four-jet production in $e^+e^-$ annihilation is able to determine the
colour factors with a precision of $\sim 20\%$.
Additional constraints may be obtained from the ratio of the
multiplicity of gluon-induced jets to that of quark-induced jets.
This has been done by DELPHI, the result is shown in figure
\ref{Fig:LEPCFSum1} \cite{Ref:DELPHIQGNch}. 
Further constraints may be obtained from the analysis of the
evolution of the means of event shape variables with energy.
This has been done using a renormalisation group invariant analysis by
DELPHI \cite{Ref:DELPHIRGI}.
The results are plotted in figure \ref{Fig:LEPCFSum1}.
JADE data has been used in a power correction fit to event shape variables
yielding $C_A = 2.84 \pm 0.24$ and $C_F = 1.29 \pm 0.18$ 
\cite{Ref:JADEPCFit}.
Overall, the data is consistent with the QCD prediction and verify
that ${\rm SU}(3)_{\rm colour}$ is the gauge group underlying QCD.

Exploiting the exquisite sensitivity of the four-jet rate to
$\alpha_{\rm S}$, ALEPH has set the colour factors to the values
predicted by QCD and used the four-jet rate to make a precise
determination of $\alpha_{\rm S}(M_Z)$ \cite{Ref:ALEPH4Jet}.
The four-jet rate as a function of a jet separation parameter is used
together with an $O(\alpha_{\rm S}^3)$ calculation that includes a
resummation of the large logarithms at next-to-leading order accuracy.
The value of $\alpha_{\rm S}(M_Z)$ obtained is:
\begin{eqnarray}
  \alpha_{\rm S} (M_Z) & = & 
  0.1170 \pm 0.0001 \pm 0.0013,
  \nonumber
\end{eqnarray}
where the first error is the statistical uncertainty and the second
error contains both the experimental systematic and the
theoretical uncertainties.
This is a determination of $\alpha_{\rm S}$ with a precision $\sim 1\%$,
a remarkable achievement.
\begin{figure}
  \begin{center}
    \includegraphics[width=0.8\columnwidth]{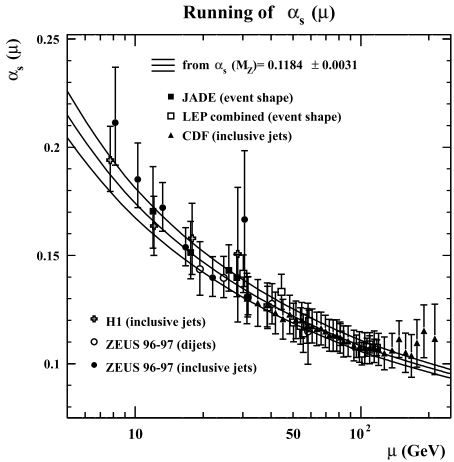}
  \end{center}
  \vspace{-0.8cm}
  \caption{
    Compilation of measurements of the running of 
    $\alpha_{\rm S}(\mu)$ with scale $\mu$. 
    Data are shown from: the combination of the final results using
    event shapes from the four LEP collaborations (open squares);
    the results of a re-analysis of event shape data from the JADE 
    experiment (solid squares);
    jet production in $ep$ collisions at HERA (solid and open crosses and
    solid circles);
    results obtained by the CDF collaboration using the inclusive 
    jet cross section in $p \bar{p}$ collisions (solid triangles).
  }
  \label{Fig:RunAlpS}
\end{figure}

The distribution of event-shape variables, such as thrust, have long
been used in the study of $e^+e^- \rightarrow {\rm hadrons}$ to
determine $\alpha_{\rm S}$.
QCD calculations at $O(\alpha_{\rm S}^2)$ matched to resummations of
the leading logarithms at NLL accuracy have been used to fit event
shape distributions by each of the LEP collaborations at each
centre-of-mass (cms) energy \cite{REF:LEPEvtShpAlfs}.
For ICHEP2002, the LEP collaborations have updated these results to
ensure a uniform treatment of the data.
The measurements of the individual experiments at each cms energy
have been combined by the LEP QCD working group, the results are shown
in figure \ref{Fig:RunAlpS}.
The data clearly exhibit the expected running of the strong coupling.
The value of $\alpha_{\rm S}$ obtained by combining all the data
at each cms energy is $\alpha_{\rm S}(M_Z) = 0.1198 \pm 0.0048$.
This result represents the last major update on the value of 
$\alpha_{\rm S}$ from event shapes at LEP.
A similar analysis has been performed using JADE data for cms
energies in the range 14~GeV to 44~GeV \cite{Ref:JADERunAlf}.
These results are also plotted in figure \ref{Fig:RunAlpS}.
The value of $\alpha_{\rm S}$ obtained by combining all 
the JADE data is $\alpha_{\rm S}(M_Z) = 0.1194^{+0.0082}_{-0.0068}$.

\begin{figure}
  \begin{center}
    \includegraphics[width=0.7\columnwidth]{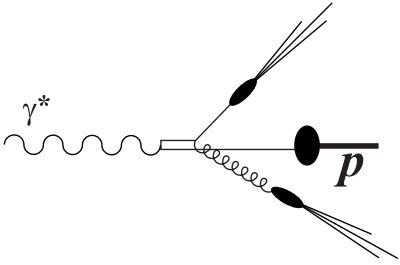}
  \end{center}
  \vspace{-0.8cm}
  \caption{
    Schematic diagram of the collision between a virtual photon and a
    quark in the Breit frame.
    The momentum of the quark is reversed by the collision.
  }
  \label{Fig:EPBreit}
\end{figure}

\begin{figure}
  \begin{center}
    \includegraphics[width=0.7\columnwidth]{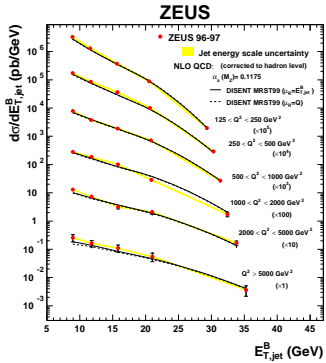}
  \end{center}
  \vspace{-0.8cm}
  \caption{
    The differential cross section $d\sigma / dE_{T,{\rm jet}}^{\rm B}$ 
    for inclusive jet production with $E_{T,{\rm jet}}^{\rm B} > 8$~GeV
    and $-2 < \eta_{\rm jet}^{\rm B} < 1.8$ in different regions of
    $Q^2$ (filled dots). 
    Each cross section has been multiplied by the scale factor
    indicated in brackets to aid visibility.
    The inner error bars show the statistical uncertainty.
    The outer error bars show the statistical and systematic
    uncertainties, not associated with the jet energy scale, added in
    quadrature. 
    The shaded band displays the uncertainty due to the jet energy
    scale. 
    The NLO QCD calculations, corrected for hadronisation effects are
    shown as the solid lines.
  }
  \label{Fig:JetETBreit}
\end{figure}

\begin{figure}
  \begin{center}
    \includegraphics[width=0.6\columnwidth]{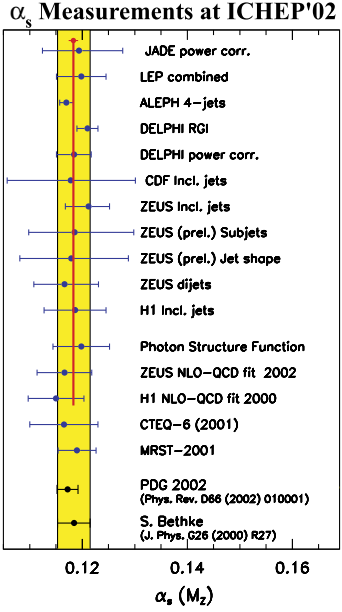}
  \end{center}
  \vspace{-0.8cm}
  \caption{
    Measurements of $\alpha_{\rm S}(M_Z)$ reported at this conference
    compared to the world average.
    The result of taking the weighted mean of the new results is also 
    indicated.
    The errors were treated as uncorrelated in this calculation.
    The long, solid, vertical line attached to the point representing
    the weighted mean shows the measurements that entered this
    calculation.
  }
  \label{Fig:AlphaSSum}
\end{figure}

The ZEUS collaboration has measured the inclusive jet cross section 
and used it to verify the running of $\alpha_{\rm S}$ and to determine
$\alpha_{\rm S}(M_Z)$ \cite{Ref:ZEUSInclJet}.
Jets were reconstructed in the Breit frame.
The Breit frame is defined such that
the current (virtual photon or $Z$ boson) is collinear with the
initial state quark, the momentum of the quark being reversed by the
collision (see figure \ref{Fig:EPBreit}).
The advantage of this choice is that the transverse energy of the
jets in the current region of the Breit frame, $E_{T,{\rm jet}}^{\rm B}$,
arises from gluon radiation.
The measured differential cross section 
$d\sigma / dE_{T,{\rm jet}}^{\rm B}$
is shown in figure \ref{Fig:JetETBreit} in bins of the virtuality of
the exchanged boson, $Q^2$.
The excellent description of the data afforded by NLO QCD allows a fit
for $\alpha_{\rm S}$ to be made.
The result, using $E_{T,{\rm jet}}^{\rm B}$ as the scale, is shown in
figure \ref{Fig:RunAlpS}.
The data exhibit the expected running and are well described by the
NLO QCD calculation.
Combining the results for $\mu = M_Z$ yields
$\alpha_{\rm S} = 0.1212 \pm 0.0017{\rm (stat.)}
^{+0.0023}_{-0.0031} {\rm (sys.)}^{+0.0028}_{-0.0027} {\rm (theory)}$.
This result is in good agreement with the world average and other
determinations of $\alpha_{\rm S}$ presented at this conference (see
figure \ref{Fig:AlphaSSum}).

CDF has measured the cross section for inclusive jet production in
$p\bar{p}$ collisions, $d\sigma / dE_{T,{\rm jet}}$, with 
$E_{T,{\rm jet}}$ measured in the laboratory frame
\cite{Ref:CDFInclJet}. 
For $E_{T,{\rm jet}} \lesssim 250$~GeV NLO QCD may be used with standard
parton density functions (PDFs) to give a good description of the data.
With the PDFs available when the data was published the data for
$E_{T,{\rm jet}} \gtrsim 250$~GeV was in excess of the NLO QCD
expectation \cite{Ref:CDFJetPap}.
The data for $E_{T,{\rm jet}} \lesssim 250$~GeV have been fitted
to extract $\alpha_{\rm S}$ using $\mu = E_{T,{\rm jet}}$.
The results are shown in figure \ref{Fig:RunAlpS}.
A value of
$\alpha_{\rm S}(M_Z) = 0.1178 \pm 0.0001{\rm (stat.)}
^{+0.0081}_{-0.0095} {\rm (sys.)}^{+0.0092}_{-0.0075} {\rm (theory)}$
is obtained by combining all these results.
The level of agreement between the inclusive jet data and NLO QCD 
is discussed in more detail in section \ref{SubSect:D0Jets}.

The measurements of $\alpha_{\rm S}(M_Z)$ reported at this conference are
shown in figure \ref{Fig:AlphaSSum}.
The new measurements, which have been obtained using a wide variety of
independent processes and techniques, are consistent both with
one-another and with the world average \cite{Ref:WorldAverage}.
The most precise of the new measurements is that obtained from the
4-jet rate in $e^+ e^-$ annihilation by the ALEPH collaboration.
Taking the weighted mean of all the new measurements, on the assumption
that the errors are uncorrelated, yields the result
$\alpha_{\rm S}(M_Z) = 0.1183 \pm 0.0009$.
The precision of this result is striking and motivates a more careful
analysis to update the world average.

\section{Proton and photon structure}
\label{Sect:ProtPhotStrct}

\begin{figure}
  \begin{center}
    \includegraphics[width=0.7\columnwidth]{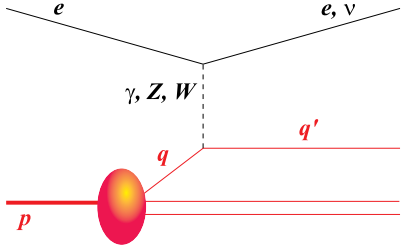}
  \end{center}
  \vspace{-0.8cm}
  \caption{
    Feynman diagram for lepton-proton deep inelastic scattering.
  }
  \label{Fig:FDDIS}
\end{figure}

\subsection{Deep inelastic $ep$ scattering}
\label{SubSect:LeptProtDIS}

The generic Feynman diagram for lepton-proton deep inelastic 
scattering (DIS) is shown in figure \ref{Fig:FDDIS}.
At HERA the incoming lepton is either an electron or a positron.
$Q^2$ is the virtuality of the exchanged boson squared and the
fraction of the four-momentum of the proton carried by the struck
quark is $x$.
The double differential cross section for the neutral current (NC)
process $e^\pm p \rightarrow e^\pm X$ may be written
\begin{eqnarray}
  \frac{d^2\sigma^{\rm NC}_{e^\pm p}}{dxdQ^2} & = & 
  \frac{2 \pi \alpha^2}{x Q^4}                     \nonumber     \\
  & &
  \left[
    Y_+ F_2^{\rm NC} \mp Y_- xF_3^{\rm NC} - y^2 F_L^{\rm NC}
  \right],
  \label{Eq:NCBorn}
\end{eqnarray}
where $\alpha$ denotes the fine structure constant, 
$Y_\pm = 1 \pm (1 - y)^2$ and $y=Q^2/xs$.
$F_L^{\rm NC}$, the longitudinal structure function, is zero at LO in
QCD, whereas the structure functions $F_2^{\rm NC}$ and $xF_3^{\rm
NC}$ can be expressed as products of electroweak couplings and
PDFs as follows:
\begin{eqnarray}
  F_2^{\rm NC} & = & x \Sigma_f 
  \left[ 
    A_f \left( q_f + \bar{q}_f \right)
  \right];                                            \nonumber     \\
 xF_3^{\rm NC} & = & x \Sigma_f 
  \left[ 
    B_f \left( q_f - \bar{q}_f \right)
  \right];                                            \nonumber 
\end{eqnarray}
where $xq_f(x,Q^2)$ are the quark and $x\bar{q}(x,Q^2)$ the
anti-quark PDFs, the functions $A_f$ and $B_f$ contain the electroweak
couplings and $f$ runs over the five quark flavours $u,...,b$.
$F_2^{\rm NC}$ may be corrected for the $Z$-exchange contribution to yield
$F_2^{\rm em}$ - the purely electromagnetic contribution to $F_2^{\rm
NC}$.
\begin{figure}
  \begin{center}
    \includegraphics[width=0.7\columnwidth]{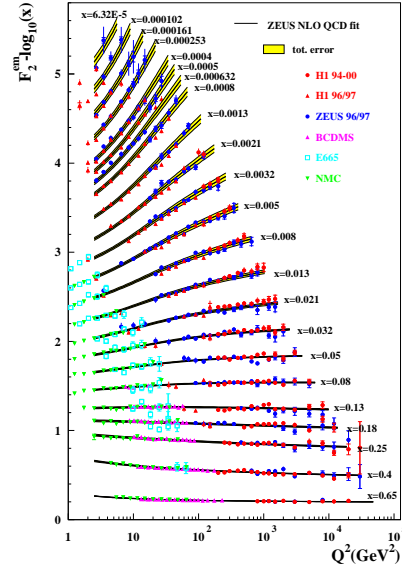}
  \end{center}
  \vspace{-0.8cm}
  \caption{
    The structure function $F_2^{\rm em}$ as a function of $Q^2$ 
    for various values of $x$. 
    The most recent data from the H1 and ZEUS collaborations are 
    shown together the measurements made by various fixed target
    experiments.
    The solid line shows the result of the ZEUS NLO QCD fit.
    The shaded band shows the uncertainty of the fit taking into
    account the statistical, correlated and uncorrelated systematic
    uncertainties.
  }
  \label{Fig:HERAF2EM}
\end{figure}
The double differential cross section for the charged current (CC)
process $e^\pm p \rightarrow \nu_e (\bar{\nu}_e) X$ may be written
\begin{eqnarray}
  \frac{d^2\sigma_{e^\pm p}^{\rm CC}}{dxdQ^2} & = &
  \frac{G_F^2}{4 \pi x} \frac{M_W^4}{\left( Q^2 + M_W^2 \right)^2}
                                                              \nonumber \\
  & &
  \left[
    Y_+ F_{2}^{\rm CC} \mp Y_- xF_{3}^{\rm CC} - y^2 F_{L}^{\rm CC}
  \right] ,
  \label{Eq:CCBorn}
\end{eqnarray}
where $G_F$ is the Fermi constant and $M_W$ the mass of the $W$ boson. 
At LO, $F_{L}^{\rm CC} = 0$ and $F_{2}^{\rm CC}$ and
$xF_{3}^{\rm CC}$ can be written in terms of the quark PDFs, for
example:
\begin{eqnarray}
  F_{2,e^+ p}^{\rm CC} & = & x
  \left[
    d  + s + \bar{u} + \bar{c}
  \right];                                                 \nonumber \\
 xF_{3,e^+ p}^{\rm CC} & = &  x
  \left[
    d  + s - \bar{u} - \bar{c}
  \right].                                                 \nonumber
\end{eqnarray}
The expressions for $e^-p$ CC DIS may be obtained by replacing quark
densities with anti-quark densities and vice versa.
The following paragraphs review recent progress in the measurement
and interpretation of neutral and charged current deep inelastic
scattering. 

\subsection{Determination of PDFs and $\boldmath{\alpha_{\rm S}}$}
\label{SubSect:DISPDFSALF}

\begin{figure}
  \begin{center}
    \includegraphics[width=0.6\columnwidth]{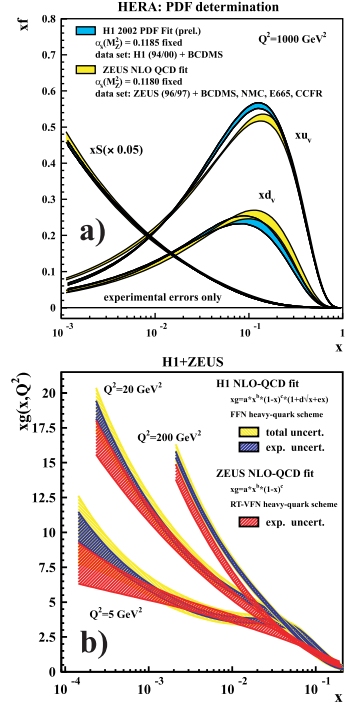}
  \end{center}
  \vspace{-0.8cm}
  \caption{
    Parton density functions extracted from the ZEUS and H1 NLO QCD
    fits.
    a) The quark PDFs plotted as a function of $x$ for
    $Q^2=1\,000$~GeV$^2$. 
    b) The gluon PDF plotted as a function of $x$ for
    $Q^2=5, 20$ and $200$~GeV$^2$. 
    The bands show the uncertainty estimated in the fit including
    the statistical, correlated and uncorrelated systematic
    uncertainties.
  }
  \label{Fig:HERAPDFs}
\end{figure}
Figure \ref{Fig:HERAF2EM} shows $F_2^{\rm em}$ as a function of $Q^2$
for several values of $x$ 
\cite{Ref:H1Pos1,Ref:H1Pos2,Ref:ZF2em,Ref:FixedTarget}.
For $x \sim 0.1$ the data show little dependence on the scale $Q^2$.
At lower $x$ values the data show a clear rise with $Q^2$.
For $Q^2$ sufficiently large, the DGLAP formalism may be used to
calculate the evolution of $F_2^{\rm em}$ with $Q^2$ by solving the 
evolution equation \cite{Ref:DGLAP}:
\begin{equation}
  \frac{dF_2^{\rm em}}{d \ln Q^2} = 
  \frac{\alpha_{\rm S}}{2 \pi} 
  \left[
    \left(
      P_{qq} \otimes F_2^{\rm em}
    \right)
    +
    \left(
      P_{qg} \otimes xG
    \right)
  \right] ,
  \label{Eq:DGLAP}
\end{equation}
where $xG(x, Q^2)$ is the gluon PDF, $P_{qq}$ and $P_{qg}$ are the
quark and gluon splitting functions and the symbol $\otimes$
represents a convolution integral.
Equation \ref{Eq:DGLAP} may be used to fit the data for both the PDFs 
and $\alpha_{\rm S}$.
NLO DGLAP fits are now routinely performed by experimental and
theoretical groups 
\cite{Ref:CTEQ6,Ref:MRST2001,Ref:H1NLOQCDFits,Ref:ZNLOQCDFit}.
Over the past year or so particular emphasis has been placed on a
full treatment of the correlated experimental systematic uncertainties.
Both the ZEUS and the H1 collaborations have performed such fits.
As an example, figure \ref{Fig:HERAF2EM} shows the ZEUS NLO QCD fit
together with the full error band which includes the contributions of
the statistical, the correlated and the uncorrelated systematic
uncertainties.
The fit describes the data well.
The error band has the tendency to grow as $x$ falls (for example at
$Q^2=100$~GeV$^2$ the fractional error at $x=0.08$ is $\sim 1\%$
while at $x=0.008$ it is $\sim 2.5\%$).

Figure \ref{Fig:HERAPDFs}a shows the quark PDFs that result from the
ZEUS and H1 fits evaluated at $Q^2=1\,000$~GeV$^2$ as a function of $x$.
The sea quark PDF, $x\Sigma$ 
($=2x\bar{u}+2x\bar{d}+xs+x\bar{s}+xc+x\bar{c}+xb+x\bar{b}$),
extracted in the two analyses agree well within errors. 
The valence quark PDFs show reasonable agreement, but differ by
between 5\% and 10\%.
The ZEUS and H1 analyses differ in the choice of data used as input to
the fit, the parameterisation chosen for the PDFs at the starting
scale and the treatment of charm and beauty quarks close to
threshold.
When the differences between the analyses are taken into account the
level of agreement between the PDFs seems reasonable.
It is interesting to note that, within a particular analysis, the
$u$-valence PDF is known with a precision of $\sim 3\%$, the
$d$-valence PDF is known to $\sim 10\%$ while the sea-quark PDF is
known to $\sim 5-10\%$.
The gluon PDF contributes to DIS only at NLO.
Therefore, the fits exhibit a strong correlation between 
$\alpha_{\rm S}$ and the parameters of $xG$.
Figure \ref{Fig:HERAPDFs}b shows the gluon PDF obtained by the ZEUS
and H1 collaborations at several values of $Q^2$ as a function of $x$.
The agreement between the analyses is reasonable, taking into account
the different choices made in setting up the fits.
The gluon density exhibits a strong rise towards low $x$.
The values of $\alpha_{\rm S}$ extracted by the collaborations from
these fits are: 
$\alpha_{\rm S}(M_Z) = 0.1150 \pm 0.0053$ (H1) \cite{Ref:H1NLOQCDFits};.
$\alpha_{\rm S}(M_Z) = 0.1166 \pm 0.0052$ (ZEUS) \cite{Ref:ZNLOQCDFit}.
The errors quoted include statistical, uncorrelated and correlated
contributions to the systematic uncertainty as well as theoretical and
model errors.
Space does not permit a full description of the error analysis. 
The CTEQ and MRST groups have also performed such analyses and have
obtained the following values:
$\alpha_{\rm S}(M_Z) = 0.1165 \pm 0.0065$ (CTEQ) \cite{Ref:CTEQ6};
$\alpha_{\rm S}(M_Z) = 0.1190 \pm 0.0036$ (MRST) \cite{Ref:MRST2001}.
The error quoted by the MRST group takes account of experimental and
theoretical uncertainties. 
The experimental contribution is determined by allowing a change in
the fit $\chi^2$ of 20, while the theoretical error is estimated by
varying the theoretical treatment to include an approximate treatment
of next-to-NLO terms, resummation of $\ln(1/x)$ or $\ln(1-x)$ terms.
The CTEQ group quotes only the experimental uncertainty on 
$\alpha_{\rm S}$, arguing that the strong correlation between 
$\alpha_{\rm S}$ and the functional form of $xG$ makes this method of
determining $\alpha_{\rm S}$ uncompetitive.
The experimental uncertainty is determined by allowing a $\chi^2$
change of 100.
This, comparatively large, value is determined by a careful study of
the variation of fit parameters allowed by the data.
Overall, the values of $\alpha_{\rm S}$ obtained from NLO QCD fits are
in good agreement.
The precision of the data and of the theoretical analysis has now
reached a stage where details of model and theoretical assumptions can
be verified.

\subsection{Search for low $\boldmath x$, low $\boldmath Q^2$ 
            limit of validity of DGLAP evolution}
\label{SubSect:DGLAPLIMIT}

The analysis of the data presented above depends on the assumption
that $Q^2$ `is large enough' for the DGLAP formalism to be applied.
A large $Q^2$ is required so that the partons may be treated as
independent non-interacting particles and so that $\alpha_{\rm S}(Q)$
is sufficiently small to allow QCD at NLO to be applied.
The $x$ values to which the fit is sensitive should be sufficiently
large that terms proportional to powers of $\alpha_{\rm S} \ln (1/x)$
are not important.
It is therefore of interest to investigate the extent to which the
DGLAP formalism describes the data at low $Q^2$ and low $x$.

\begin{figure}
  \begin{center}
    \includegraphics[width=1.0\columnwidth]{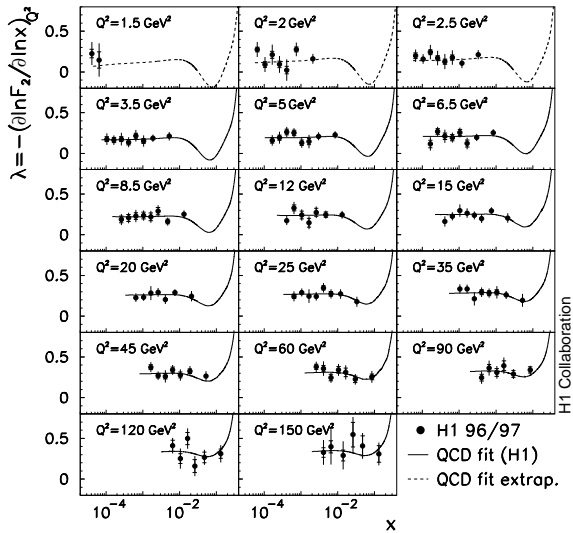}
  \end{center}
  \vspace{-0.8cm}
  \caption{
    H1 measurement of the dependence of the parameter $\lambda$ on $x$ as
    a function of $Q^2$.
    The inner error bars represent the statistical uncertainty while
    the full error bars include the systematic uncertainty added in
    quadrature. 
    The solid curves show the results of the H1 NLO QCD fit.
    The dashed curves show the extrapolation of the fit below
    $Q^2=3.5$~GeV$^2$.
  }
  \label{Fig:H1LamVsX}
\end{figure}
\begin{figure}
  \begin{center}
    \includegraphics[width=0.8\columnwidth]{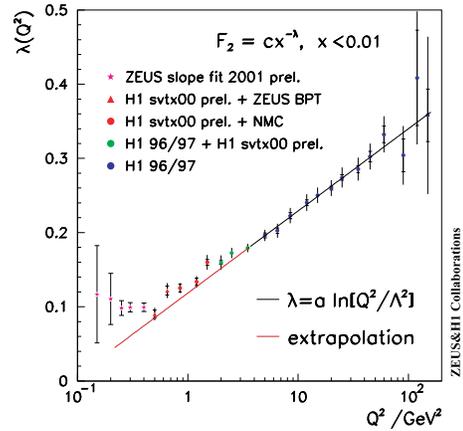}
  \end{center}
  \vspace{-0.8cm}
  \caption{
    The $Q^2$ dependence of the parameter $\lambda$.
    The data (solid points) were obtained from fits to ZEUS and H1
    structure function data.
    The inner error bars represent the statistical uncertainty while
    the full error bars include the systematic uncertainty added in
    quadrature. 
    The solid line shows the result of performing a straight line fit
    to the data for $Q^2 > 3.5$~GeV$^2$.
  }
  \label{Fig:ZH1LamPlot}
\end{figure}

One approach to this is to parameterise $F_2^{\rm em}$ at a particular
$Q^2$ as a power of $x$ as follows:
\begin{eqnarray}
  F_2^{\rm em} = c x^{-\lambda},
  \nonumber
\end{eqnarray}
where $c$ and $\lambda$ are positive parameters.
Neighbouring data points at a particular $Q^2$ may be combined to
estimate the slope parameter $\lambda$.
This has been done by the H1 collaboration and the results
are shown in figure \ref{Fig:H1LamVsX} \cite{Ref:H1Lambda}.
For $x \gtrsim 10^{-2}$ the contribution of the valence quarks to 
$F_2^{\rm em}$ is large and the ansatz $F_2^{\rm em} \propto x^{-\lambda}$
does not hold.
However, for $x \lesssim 10^{-2}$ the data is well described by a single value
of $\lambda$, independent of $x$ over the range of $x$ to which the
experiment is sensitive.
Hence, at each value of $Q^2$, $F_2^{\rm em}$ may be characterised by
a particular value of $\lambda$.
The dependence of $\lambda$ on $Q^2$ is shown in figure
\ref{Fig:ZH1LamPlot} \cite{Ref:H1Lambda,Ref:ZEUSLambda}.
For $Q^2 \gtrsim 4$~GeV$^2$, where the DGLAP formalism gives a good
description of the data, the dependence of $\lambda$ on $Q^2$ is, to a
good approximation, linear.
However, for $Q^2 \lesssim 1$ ~GeV$^2$ the data lie above the extrapolation of
this linear dependence.

It is intriguing to note that while no significant dependence of
$\lambda$ on $x$ for $x \lesssim 10^{-2}$ has been observed, there is a change
in the dependence of $\lambda$ on $Q^2$ at $Q^2 \sim 1$~GeV$^2$.
The explanation for this change of behaviour is the subject of intense
theoretical speculation.

\subsection{Deep inelastic scattering at high $Q^2$}
\label{SubSect:HighQ2DIS}

\begin{figure}
  \begin{center}
    \includegraphics[width=0.8\columnwidth]{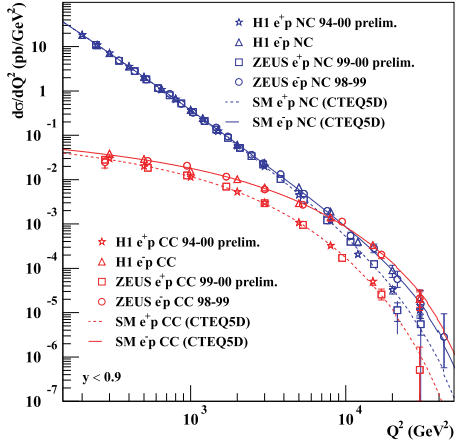}
  \end{center}
  \vspace{-0.8cm}
  \caption{
    Measurement of the differential cross section $d \sigma / d Q^2$
    for neutral and charged current deep inelastic scattering by the
    ZEUS and H1 collaborations.
  }
  \label{Fig:HERADSDQ2}
\end{figure}

Figure \ref{Fig:HERADSDQ2} shows the differential cross section for NC and
CC $e^\pm p$ scattering at high $Q^2$
\cite{Ref:H1Pos1,Ref:H1Pos2,Ref:H1Ele,Ref:ZPos1NC,Ref:ZPos1CC,Ref:ZEleNC,Ref:ZEleCC,Ref:ZPos2}.
The NC cross section falls by 3--4 orders of magnitude between
$Q^2 = 200$~GeV$^2$ and $Q^2 = 2\,000$~GeV$^2$, exhibiting the $1/Q^4$
dependence of the dominant single-photon exchange contribution.
For $Q^2 \gtrsim 5\,000$~GeV$^2$ the $e^-p$ NC DIS cross section lies above
that for $e^+p$ NC DIS reflecting the fact that the $\gamma Z$
interference contribution is constructive in the case of $e^-p$
scattering and destructive for $e^+p$.
The $e^-p$ CC DIS cross section is always larger than that for $e^+p$
CC DIS because the $u$-quark density, which gives the dominant
contribution, is larger than that of the
$d$-quark and, in the case of $e^+p$ CC DIS, the $d$-quark
contribution is suppressed by $(1-y)^2$ (see equation \ref{Eq:RedCC}).
The SM, evaluated using standard PDFs evolved in the DGLAP formalism,
gives a good description of the data.
This motivates the use of NC and CC data at high $Q^2$ to study
proton structure at high $Q^2$ and high $x$.

\begin{figure}
  \begin{center}
    \includegraphics[width=1.0\columnwidth]{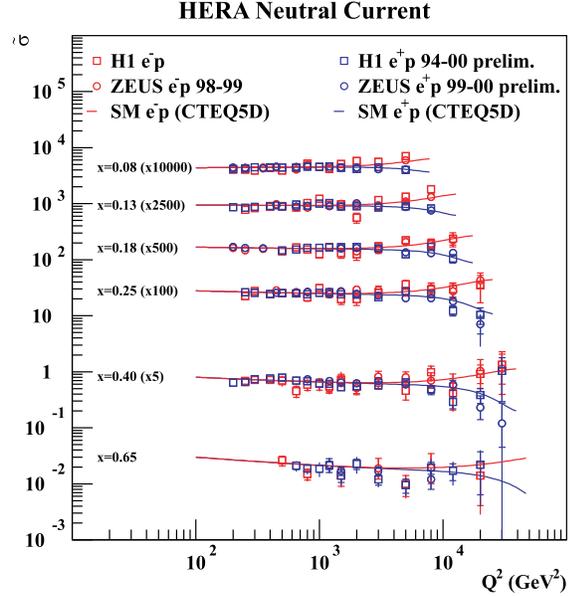}
  \end{center}
  \vspace{-0.8cm}
  \caption{
    The reduced cross section $\tilde{\sigma}^{\rm NC}_{e^\pm p}$
    plotted as a function of $Q^2$ for various values of $x$.
    The points represent the data while the solid lines represent the 
    expectation of the Standard Model evaluated with the ZEUS NLO
    QCD fit.
  }
  \label{Fig:HERARedNC}
\end{figure}

\begin{figure}
  \begin{center}
    \includegraphics[width=1.0\columnwidth]{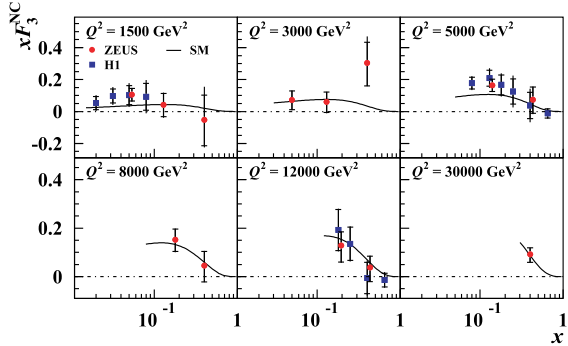}
  \end{center}
  \vspace{-0.8cm}
  \caption{
    The structure function $xF_3^{\rm NC}$ plotted as a function of
    $x$ for various values of $Q^2$. 
    The points represent the data while the solid lines represent the 
    expectation of the Standard Model evaluated with the ZEUS NLO
    QCD fit.
  }
  \label{Fig:HERAxF3}
\end{figure}

In order to exhibit the dependence of the DIS cross sections upon the
structure functions it is convenient to divide out the dependence on
the electroweak couplings and the boson propagator.
The scaled cross sections, referred to as reduced cross sections, for
NC and CC DIS are defined as follows:
\begin{eqnarray}
  \tilde{\sigma}^{\rm NC}_{e^\pm p} & = &
  \left[
    \frac{2 \pi \alpha^2}{x Q^4} 
  \right]^{-1}
  \frac{d^2\sigma^{\rm NC}_{e^\pm p}}{dxdQ^2};    \nonumber    \\
  \tilde{\sigma}^{\rm CC}_{e^\pm p} & = &
  \left[
    \frac{G_F^2}{4 \pi x} \frac{M_W^4}{\left( Q^2 + M_W^2 \right)^2}
  \right]^{-1}
  \frac{d^2\sigma_{e^\pm p}^{\rm CC}}{dxdQ^2}.    \nonumber
\end{eqnarray}
Figure \ref{Fig:HERARedNC} shows $\tilde{\sigma}^{\rm NC}_{e^\pm p}$ as a
function of $Q^2$ for several values of $x$.
The photon propagator dependence having been removed,
$\tilde{\sigma}^{\rm NC}$ depends only weakly on $Q^2$.
As described above,  the effect of $Z$-boson exchange is to cause 
$\tilde{\sigma}^{\rm NC}_{e^-p}$ to lie above 
$\tilde{\sigma}^{\rm NC}_{e^+p}$ for $Q^2 \gtrsim 5\,000$~GeV$^2$.
The difference between $\tilde{\sigma}^{\rm NC}_{e^-p}$ and
$\tilde{\sigma}^{\rm NC}_{e^+p}$ can be used to determine 
$xF_3^{\rm NC}$ (see equation \ref{Eq:NCBorn}).
This has been done by the H1 and ZEUS collaborations and the final
results for pre-upgrade HERA running are shown in figure
\ref{Fig:HERAxF3} \cite{Ref:H1Ele,Ref:ZEleNC}.
Since $xF_3^{\rm NC} \propto \Sigma_q ( q - \bar{q})$, $xF_3^{\rm NC}$
gives a measure of the valence quark PDF.
The data are well described by the SM expression evaluated with
standard PDFs.
The large data sets that will be provided in the near future at HERA
will allow a precise determination of $xF_3^{\rm NC}$ to be made
yielding an important constraint on the valence quark PDFs.

\begin{figure}
  \begin{center}
    \includegraphics[width=1.0\columnwidth]{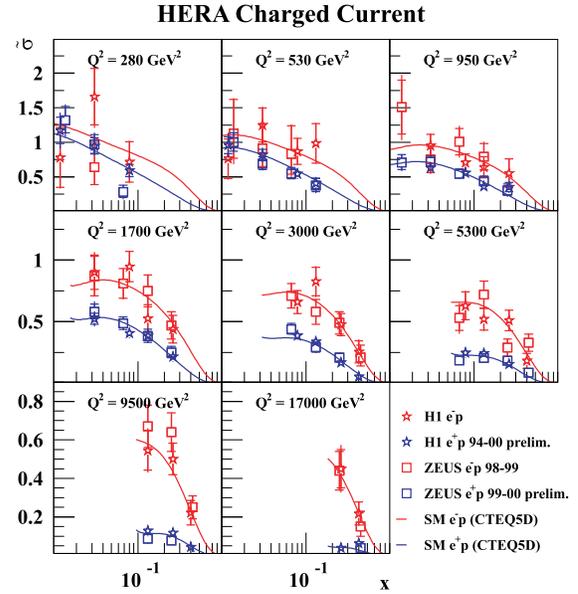}
  \end{center}
  \vspace{-0.8cm}
  \caption{
    The reduced cross section $\tilde{\sigma}^{\rm CC}_{e^\pm p}$
    plotted as a function of $Q^2$ for various values of $x$.
    The points represent the data while the solid lines represent the 
    expectation of the Standard Model evaluated with the ZEUS NLO
    QCD fit.
  }
  \label{Fig:HERARedCC}
\end{figure}

At LO in $\alpha_{\rm S}$, $\tilde{\sigma}^{\rm CC}_{e^\pm p}$ can be
written 
\begin{equation}
  \tilde{\sigma}^{\rm CC}_{e^+p} = d + s +
  (1-y)^2 \left( \bar{u} + \bar{c} \right) .
  \label{Eq:RedCC}
\end{equation}
The expression for $e^-p$ CC DIS may be obtained by replacing
$u$-type quark densities with $d$-type quark densities and vice
versa. 
A measurement of the CC cross section at high $x$ can therefore be
used to constrain the valence $u$- ($e^-p$) and valence $d$-quark
($e^+p$) PDFs. 
Figure \ref{Fig:HERARedCC} shows $\tilde{\sigma}^{\rm CC}_{e^\pm p}$ as a
function of $x$ at several values of $Q^2$.
The data are well described by the SM.
Also shown, are the contributions of the $u$-type quarks in $e^-p$ CC
DIS and the $d$-type quarks in $e^+p$.
The statistical precision of the data is not yet sufficient to allow a
precise determination of the $u$- and $d$-quark PDFs.
The large data sets expected to be obtained at HERA II will allow
precise measurements of the $e^\pm p$ CC DIS cross sections to be
made. 
This will yield important constraints on the $u$- and $d$-quark PDFs.

\subsection{Impact of D0 inclusive jet data}
\label{SubSect:D0Jets}

\begin{figure}
  \includegraphics[width=1.0\columnwidth]{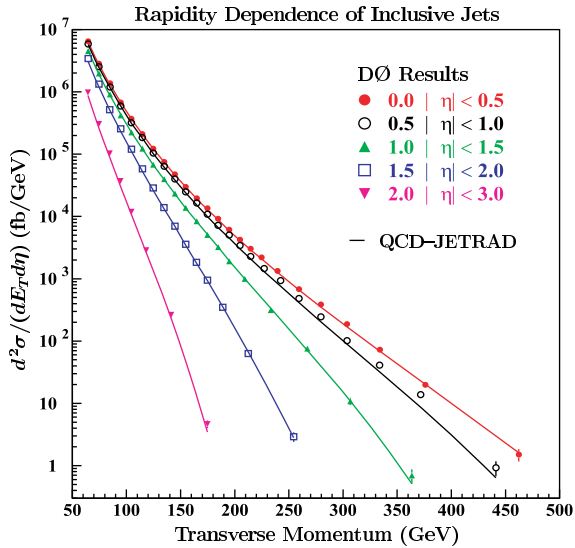}
  \vspace{-0.8cm}
  \caption{
    The double differential inclusive jet cross section for the
    process $p \bar{p} \rightarrow {\rm Jet}+X$ measured by the D0
    collaboration (points) versus jet transverse energy in intervals
    of jet pseudorapidity.
    The solid lines show the result of a NLO QCD calculation.
  }
  \label{Fig:D0InclJets}
\end{figure}

Last year the D0 collaboration published their measurement of the
inclusive jet double differential cross section as a function of jet
transverse energy, $E_{T,{\rm jet}}$, in bins of jet pseudorapidity, 
$\eta=-\ln [ \tan (\theta_{\rm jet}/2) ]$, where 
$\theta_{\rm jet}$ is the polar angle of the jet in the D0 laboratory
frame \cite{Ref:D0InclJets}.
The data, shown in figure \ref{Fig:D0InclJets}, are well described by QCD
at NLO over the full range of $E_{T,{\rm jet}}$ and
$\eta$.

\begin{figure}
  \includegraphics[width=1.0\columnwidth]{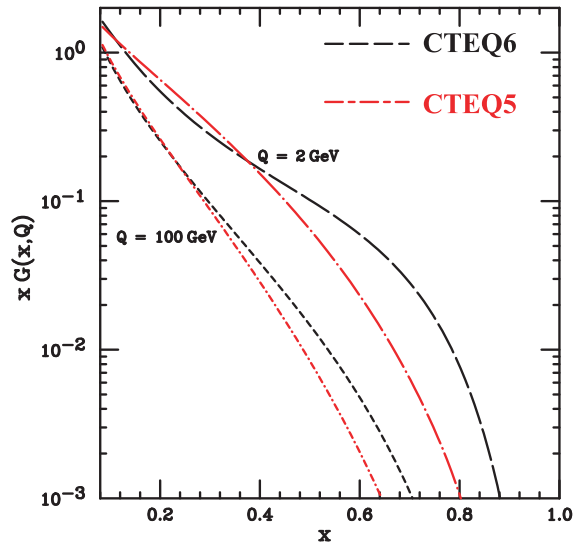}
  \vspace{-0.8cm}
  \caption{
    Comparison of the CTEQ6 (dashed) to CTEQ5 (dash-dotted) gluon
    distributions at scales of 2 and 100 GeV.
  }
  \label{Fig:CTEQGlu}
\end{figure}

The CTEQ and MRST groups have used the data in fits to determine the
proton PDFs \cite{Ref:CTEQ6,Ref:MRST2001}.
The importance of this data in constraining the gluon PDF can
be judged from figure \ref{Fig:CTEQGlu} where $xG$ is plotted as a
function of $x$ for two values of the scale $Q$.
The new result (labelled CTEQ6) is compared to the result of a fit
that did not include this data (CTEQ5) \cite{Ref:CTEQ5}.
The gluon density obtained in the new fits is much harder than that
obtained previously.
It is important to note that the new PDFs, containing the enhanced
gluon density at high $x$, give a good description of both the CDF and
the D0 jet data, including the data at high transverse energy 
($E_{T,{\rm jet}} \gtrsim 250$~GeV) in the central region 
($|\eta| < 0.5$) which were in excess of QCD at NLO evaluated with 
older PDFs.

\subsection{Remark on the precision of PDFs}
\label{SubSect:PDFPrecRem}

NLO QCD fits to data are able to determine the quark PDFs with a
precision of ~5--10\% and the gluon PDF with a precision of 10--15\%.
Progress is being made towards calculations of many of the cross
sections used in the fits at next-to-NLO (NNLO).
PDFs extracted from a fit using a partial NNLO calculation
have recently become available \cite{Ref:MRSTNNLO}.
A consistent analysis at NNLO combined with the large data sets soon
to become available from HERA and the Tevatron will allow a
significant reduction in the PDF uncertainties to be obtained before
the LHC era begins.

\subsection{Photon structure}
\label{SubSect:PhotStruc}

\begin{figure}
  \begin{center}
    \includegraphics[width=0.8\columnwidth]{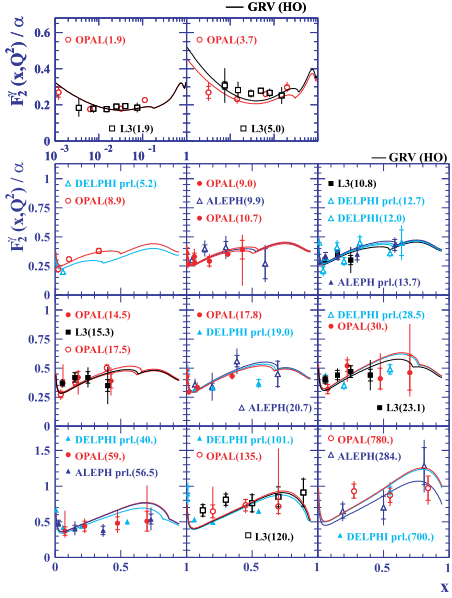}
  \end{center}
  \vspace{-0.8cm}
  \caption{
    Compilation of measurements of the structure function $F_2^\gamma$
    for real photons from $e^+e^-$ scattering as a function of $x$ for
    various values of $Q^2$.
    Note that $Q^2$ here denotes the virtuality of the photon that
    probes the structure of the quasi-real photon.
  }
  \label{Fig:F2GammaSum}
\end{figure}

The LEP collaborations have updated their measurements of the photon
structure functions and contributed new measurements at the highest
LEP energies \cite{Ref:LEPF2Gam}.
Figure \ref{Fig:F2GammaSum} shows a compilation of all measurements of
the photon structure function $F_2^\gamma$.
The data are reproduced to within $\sim 20\%$ by the NLO QCD
calculation using various parameterisations of $F_2^\gamma$.
Motivated by this level of agreement, the value of $\alpha_{\rm S}$
has been extracted from a fit to the data with the result
$\alpha_{\rm S}(M_Z) = 0.1198 \pm 0.0054$ \cite{Ref:F2GamAlf}.
The result is in good agreement with the world average and other recent
results (see figure \ref{Fig:AlphaSSum}).
It will be important to establish the insensitivity of the result to
the form of the parameterisation of the gluon density if the result is
to be included in a future world average $\alpha_{\rm S}(M_Z)$.

\section{Deep inelastic diffraction}
\label{Sect:Diffraction}

\begin{figure}
  \begin{center}
    \includegraphics[width=0.6\columnwidth]{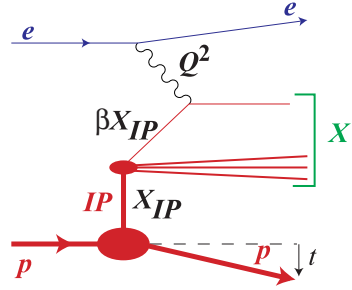}
  \end{center}
  \vspace{-0.8cm}
  \caption{
    Schematic Feynman diagram of diffraction in deep inelastic scattering.
  }
  \label{Fig:FDDiff}
\end{figure}

The sample of NC DIS events observed by the H1 and ZEUS collaborations
contains an intriguing diffractive subsample.
Events in this subsample are characterised by the presence of a proton
in the final state at small $|t|$ where $\sqrt{|t|}$ is the
four-momentum transfer between the incoming and outgoing proton. 
The subsample is large, acounting for $\sim 10\%$ of all NC DIS
events. 
Figure \ref{Fig:FDDiff} shows a schematic Feynman diagram for deep
inelastic diffraction.
The incoming proton scatters through a small angle, emerging from the
collision with an energy $\sim ( 1 - x_{\IP} )E_p$ (where $E_p$ is the
proton beam energy) and a transverse momentum kick of $\sim \sqrt{|t|}$.
The four-momentum lost by the proton is carried into the deep
inelastic scatter by a colourless partonic state often referred to as
the Pomeron.
The electroweak current then picks out a parton, generating a hadronic
system $X$, of mass $M_X$, which may be observed in the detector.
Note that the hadronic excitation produced in the deep-inelastic
scatter is not colour connected to the scattered proton.
As a result, diffractive deep inelastic scattering events are
characterised by an absense of hadronic energy flow between the proton
and the hadronic system $X$.

\begin{figure}
  \begin{center}
    \includegraphics[width=0.7\columnwidth]{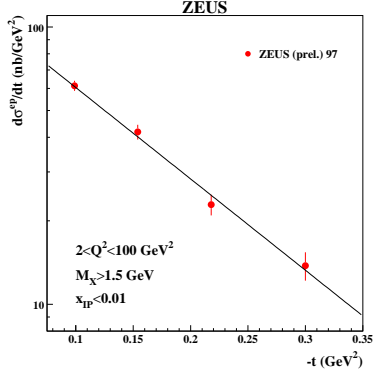}
  \end{center}
  \vspace{-0.8cm}
  \caption{
    The differential cross section $d \sigma / d |t| $ for the reaction
    $e p \rightarrow epX$ in the kinematic range indicated on the plot.
    The error bars show the statistical uncertainty.
  }
  \label{Fig:ZEUSt}
\end{figure}

Two methods have been used to select such events.
The most direct method is to tag the scattered proton.
In this case the cross section can be measured as a function of $t$.
The most recent measurement from the ZEUS collaboration is shown in
figure \ref{Fig:ZEUSt} \cite{Ref:ZEUSdsdt}.
The cross section exhibits the exponentially falling behaviour typical
of diffractive processes.
A fit to the data of the form $d\sigma / dt \propto \exp ( -b |t|)$
yields $b=7.8 \pm 0.5 ^{+0.9}_{-0.6}$~GeV$^2$.
The second method used to select diffractive events exploits the lack
of hadronic activity in the forward (proton) direction.
This selection is referred to as the large rapidity-gap selection.
It is not possible with the large rapidity-gap selection to
distinguish between events in which the proton scatters elastically
from those in which a low mass excitation is produced at the proton
vertex making it necessary to estimate the size of such a dissociative
contribution and make an appropriate correction to the data.

\begin{figure}
  \begin{center}
    \includegraphics[width=1.0\columnwidth]{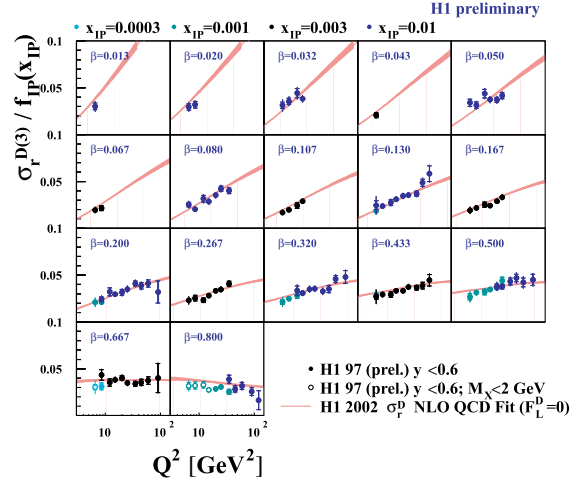}
  \end{center}
  \vspace{-0.8cm}
  \caption{
    $Q^2$ dependence of the reduced diffractive cross section scaled
    at each $x_{\IP}$ by the assumed $t$-integrated Pomeron flux.
    The data are compared with the prediction of the NLO QCD fit.
  }
  \label{Fig:H1Sig3}
\end{figure}

\begin{figure}
  \begin{center}
    \includegraphics[width=1.0\columnwidth]{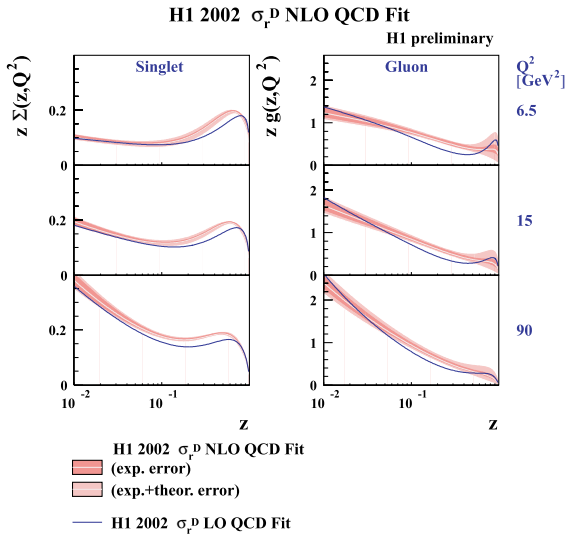}
  \end{center}
  \vspace{-0.8cm}
  \caption{
    Diffractive parton densities determined in the H1 NLO QCD fit
    to the reduced diffractive cross section.
    The left-hand side shows the singlet quark distribution.
    The right-hand side shows the gluon density.
    The light shaded band shows the total error (experimental and
    theoretical) estimated using the fit, while the inner band (more
    heavily shaded) shows the experimental (statistical and systematic)
    uncertainties.
    The result of a LO fit is shown as the solid line.
  }
  \label{Fig:H1DiffPDF}
\end{figure}

\begin{figure}
  \begin{center}
    \includegraphics[width=1.0\columnwidth]{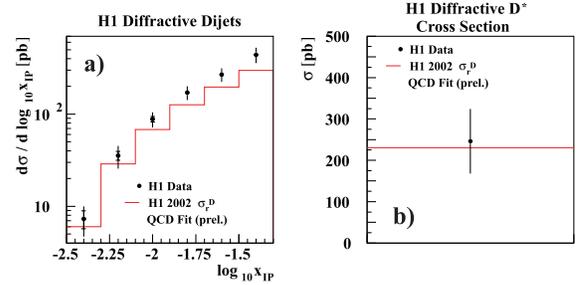}
  \end{center}
  \vspace{-0.8cm}
  \caption{
    Comparison of the H1 measurement of (a) the diffractive dijet cross 
    section and (b) the diffractive $D^*$ cross section with the result 
    of a NLO QCD calculation using the diffractive parton densities 
    extracted from the H1 NLO QCD fit to the reduced diffractive cross 
    section.
  }
  \label{Fig:H1DiffDStarDiJet}
\end{figure}

Substantial experimental and theoretical progress has been made in
recent years.
An important theoretical development was the proof that at a
particular $x_{\IP}$ and $t$, the cross section for the inclusive
diffractive process $ep \rightarrow eXp$ may be written in the form
\cite{Ref:Collins}:
\begin{eqnarray}
  \sigma^{\rm D(4)} & \propto & \Sigma_f 
  \left[
    q_f^{\rm Diff} \left(x_{\IP}, t; \beta , Q^2 \right) \otimes 
    \sigma_f \left( \beta , Q^2 \right)
  \right] ,                                                   \nonumber
\end{eqnarray}
where $q_f^{\rm Diff} \left(x_{\IP}, t; \beta, Q^2 \right)$ are diffractive 
parton densities, $\sigma_f \left( \beta , Q^2 \right)$ is the hard $eq$ deep
inelastic scattering cross section and the sum runs over all active
flavours $f$.  
A consequence of this is that a determination of the diffractive cross
section at fixed $x_{\IP}$ and $t$ as a function of $\beta$ and
$Q^2$ may be used in a NLO DGLAP analysis to determine the
diffractive PDFs.
The H1 collaboration has exploited the fact that the data are
consistent with the hypothesis of Regge factorisation to carry out
such an analysis \cite{Ref:H1PomFit}.
A reduced diffractive cross section was defined in analogy to 
$\tilde{\sigma}_{e^\pm p}^{\rm NC}$ as follows:
\begin{eqnarray}
  \tilde{\sigma}^{\rm D(4)} & = &
  \left[
    \frac{2 \pi \alpha^2}{\beta Q^4}
  \right]^{-1}
  \frac{d^4 \sigma}{dx_{\IP} dt d\beta Q^2}. \nonumber
\end{eqnarray}
Regge factorisation implies that the diffractive parton densities
do not vary in shape with $x_{\IP}$ so that 
$\tilde{\sigma}^{\rm D(4)}$ factorises into a convolution of the 
``Pomeron flux''
$f_{\IP}\left( x_{\IP}, t \right)$ and a hard scattering cross section
$\tilde{\sigma} \left( \beta, Q^2 \right)$
as follows:
\begin{eqnarray}
  \tilde{\sigma}^{\rm D(4)} & = &
  f_{\IP}\left( x_{\IP}, t \right) \otimes 
  \tilde{\sigma} \left( \beta, Q^2 \right). \nonumber
\end{eqnarray}
Integration of $\tilde{\sigma}^{\rm D(4)}$ over $t$ yields:
\begin{eqnarray}
  \tilde{\sigma}^{\rm D(3)} & = &
  f_{\IP}\left( x_{\IP} \right) \times 
  \tilde{\sigma} \left( \beta, Q^2 \right). \nonumber
\end{eqnarray}
Figure \ref{Fig:H1Sig3} shows the H1 measurement of 
$\tilde{\sigma}^{\rm D(3)}/f_{\IP}\left( x_{\IP} \right)$ plotted
as a function of $Q^2$ for several values of $\beta$.
For $\beta < 0.6$ the data rise with $Q^2$ indicating that the
diffractive PDFs are dominated by gluons.
The H1 collaboration has performed a NLO DGLAP fit the result of
which is also shown in figure \ref{Fig:H1Sig3} and gives a good
description of the data.
The diffractive PDFs extracted from the fit are shown in figure
\ref{Fig:H1DiffPDF}.
Noting that the sum of the quark PDFs ($\Sigma$) is plotted on an
expanded scale, it is clear that the diffractive gluon PDF is much
larger than that of the quarks.
An important property of the PDFs of the proton is that they are
universal.
In order to investigate to what extent the diffractive PDFs are
universal the H1 collaboration has calculated the diffractive 
dijet rate and the cross section for diffractive $D^*$ production.
The results of these calculations are compared to the H1 measurements
in figure \ref{Fig:H1DiffDStarDiJet}.
With the current precision of the data, and of the diffractive PDFs,
the calculation gives a reasonable description of the data and
therefore lends support to the notion that the diffractive PDFs are
universal.

\begin{figure}
  \begin{center}
    \includegraphics[width=1.0\columnwidth]{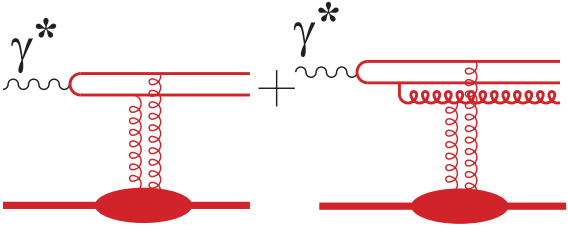}
  \end{center}
  \vspace{-0.8cm}
  \caption{
    Schematic diagram of diffractive deep inelastic scattering
    in the dipole model.
  }
  \label{Fig:DipoleModel}
\end{figure}

\begin{figure}
  \begin{center}
    \includegraphics[width=1.0\columnwidth]{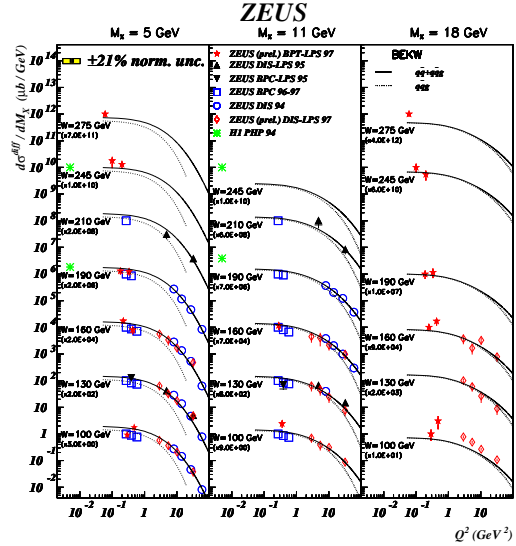}
  \end{center}
  \vspace{-0.8cm}
  \caption{
    The ZEUS measurement of the diffractive cross section 
    $d \sigma / d M_X$ plotted
    as a function of $Q^2$ for various values of $M_X$ and $Q^2$.
    The solid lines are the result of the dipole model parameterisation
    outlined in the text.
  }
  \label{Fig:ZEUSDiffMX}
\end{figure}

An alternative approach to the interpretation of diffractive DIS is to
build a model of the interaction. 
This is usually done in the framework of a dipole model in which
the interaction between the virtual photon and the proton is assumed
to arise from the exchange
of a Pomeron between the proton and a colour dipole parton system
(see figure \ref{Fig:DipoleModel}) \cite{Ref:CDM}.
The colour dipole system arises as a vacuum fluctuation in the wave
function of the virtual photon, at lowest order a $q\bar{q}$ pair,
at NLO a colour dipole $q\bar{q}g$ state.
The exchanged Pomeron is modelled as a collection of gluons in a
colour singlet state.
The simplest (lowest order) configuration is a pair of
gluons.
Parameters governing the functional form of the parameterisation of 
the cross section were determined in fits to data on inclusive 
diffraction in DIS.
The ZEUS collaboration has measured the differential cross section
$d\sigma / d M_X$ \cite{ZEUS:dsdMX}.
The data is shown at fixed photon-proton cms energy ($W$) as a function 
of $Q^2$ for several values of $M_X$ together with the predictions of 
the colour dipole model in figure \ref{Fig:ZEUSDiffMX}.
The data rise as  $Q^2$ falls, reaching a plateau for $Q^2 \sim
1$~GeV$^2$. 
The colour dipole model gives a good description of the data.
The $q \bar{q} g$ contribution dominates, particularly at low $Q^2$.

\section{Jet production: an example}
\label{Sect:JetProd}

Substantial progress has been made in the measurement of a variety of
inclusive jet topologies both at HERA and at the Tevatron.
At the same time, theoretical progress in the development of codes to
calculate at fixed order has also been made.
It is not possible to make a complete survey of all the important
contributions in the space available.
QCD at NLO is often able to describe the qualititative features of the
data.
However, it is often not possible to obtain quantitative information,
such as measurements of $\alpha_{\rm S}$ or parton densities.
The reason for this is either that the data or the theory (or both)
is too imprecise.
The complementarity of the precision achieved in jet measurements
and in the theoretical predictions at LEP is striking and has
led to precise determinations of such quantities as 
$\alpha_{\rm S}$.
I have chosen to discuss the photoproduction of dijets below as an
example of the need to strive for a similar complementarity in
experimental and theoretical precision in hadron induced processes.
Dijet photoproduction is particularly exciting because it offers
sensitivity to photon and proton PDFs, especially at high $x$, and 
sensitivity to $\alpha_{\rm S}$.

\begin{figure}
  \begin{center}
    \includegraphics[width=0.8\columnwidth]{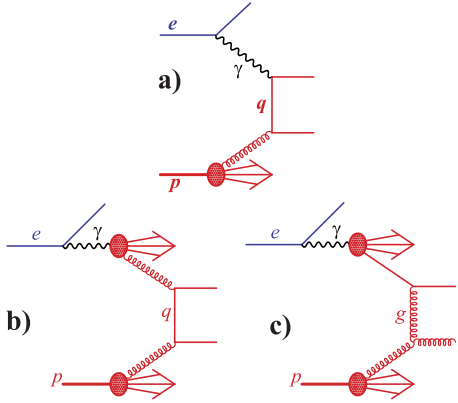}
  \end{center}
  \vspace{-0.8cm}
  \caption{
    Examples of Feynman diagrams for the dijet photoproduction of.
    (a) LO diagram contributing to the direct process.
    (b) and (c) Examples of diagrams contributing at LO to the 
    resolved process.
  }
  \label{Fig:DiJetGamPFD}
\end{figure}

The LO Feynman diagrams contributing to dijet photoproduction are shown
in figure \ref{Fig:DiJetGamPFD}.
At leading order, the direct process occurs through photon-gluon
fusion.
Two diagrams contribute, at LO, to the resolved process (figures
\ref{Fig:DiJetGamPFD}b and c).
The $qg$ scattering diagram, in which a gluon is exchanged, gives the
dominant contribution to the resolved process.
Both ZEUS and H1 have presented measurements of the cross sections for
dijet photoproduction and have shown that these measurements are well
described by NLO QCD \cite{Ref:H1PhotDiJets,Ref:ZPhotDiJets}.

If the two jet transverse energies ($E_{T,{1}}$ and 
$E_{T,{2}}$) and the jet pseudo-rapidities ($\eta_{1}$
and $\eta_{2}$) are measured, the fraction of the momentum of
the photon ($x_\gamma^{\rm Obs}$) and the fraction of the proton
momentum ($x_p^{\rm Obs}$) entering the collision can be estimated
from the following formul\ae:
\begin{eqnarray}
  x_\gamma^{\rm Obs} & = & 
  \frac{ E_{T,{1}} \exp \left( -\eta_{1} \right) + 
         E_{T,{2}} \exp \left( -\eta_{2} \right)
        }
       { 2 y E_e };                                   \nonumber   \\
  x_p^{\rm Obs} & = & 
  \frac{ E_{T,{1}} \exp \left( \eta_{1} \right) + 
         E_{T,{2}} \exp \left( \eta_{2} \right)       \nonumber
        }
       { 2 y E_e }.
\end{eqnarray}
A cut on $x_\gamma^{\rm Obs}$ can be used to obtain a sample enriched
in resolved or direct events.
A measurement of the dijet photoproduction cross section as a function
of $x_\gamma^{\rm Obs}$ ($x_p^{\rm Obs}$) is sensitive to the photon
(proton) PDFs.

\begin{figure}
  \begin{center}
    \includegraphics[width=0.8\columnwidth]{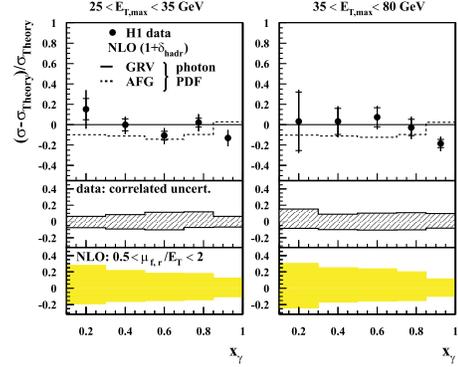}
  \end{center}
  \vspace{-0.8cm}
  \caption{
    The $x_\gamma^{\rm Obs}$ dependence of the relative difference of
    the measured dijet photoproduction cross section from the NLO QCD
    prediction.
    The inner error bars show the statistical uncertainty, while the
    outer error bars show the sum of statistical and uncorrelated
    systematic uncertainties.
    The correlated systematic errors are shown in the hatched band in
    the middle plot.
    The lower plot shows, as the shaded band, the renormalisation and
    factorisation scale uncertainties of the NLO calculation.
  }
  \label{Fig:H1DiJets}
\end{figure}

H1 has measured the dijet photoproduction cross section for 
$E_{T,{1}} > 25, E_{T,{2}} > 15$~GeV and 
$-0.5 < \eta_{1,2} < 2.5$.
The measurement is compared to the predictions of QCD at NLO in figure 
\ref{Fig:H1DiJets} for two ranges of $E_{T,{\rm max}}$ 
($= E_{T,{1}} + E_{T,{2}}$).
For $x_\gamma^{\rm Obs} < 0.8$ the data are well described by the
calculation.
The theoretical and experimental uncertainties are presented in the
figure.
It is interesting to note that the experimental statistical and
systematic errors are of comparable size and are somewhat smaller than
the QCD scale uncertainty.

\begin{figure}
  \begin{center}
    \includegraphics[width=0.8\columnwidth]{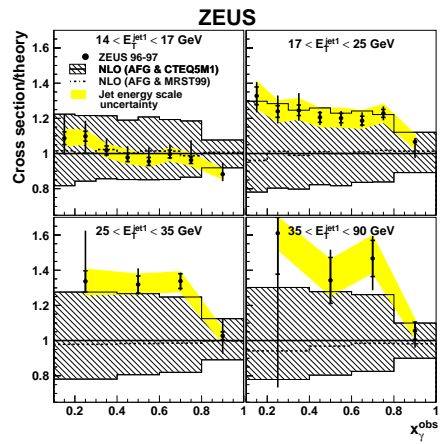}
  \end{center}
  \vspace{-0.8cm}
  \caption{
    Ratio of the ZEUS measurement of the cross section for dijet 
    photoproduction to the NLO QCD prediction as a function of 
    $x_\gamma^{\rm Obs}$ in four regions of $E_{T,1}$.
    The data are shown with statistical errors (inner bars) and statistical
    and systematic errors added in quadrature (outer bars).
    The shaded band shows the uncertainty due to that of the jet energy 
    scale.
    The theoretical uncertainty is shown as the hatched band.
  }
  \label{Fig:ZEUSDiJets}
\end{figure}

ZEUS has also measured the dijet photoproduction cross section using
the following selection: 
$E_{T,{1}} > 14, E_{T,{2}} > 12$~GeV and 
$-1 < \eta_{1,2} < 2.4$.
The measurement is compared to the NLO QCD prediction in figure
\ref{Fig:ZEUSDiJets} as a function of $x_\gamma^{\rm Obs}$ in bins of 
$E_{T,{1}}$.
The figure indicates that the data falls less steeply with 
$E_{T,{1}}$ than the NLO QCD calculation.
Again, the experimental statistical and systematic errors are of
comparable size and, in the lower $E_{T,{1}}$ bins, are
smaller than the theoretical uncertainty.

\begin{figure}
  \begin{center}
    \includegraphics[width=0.8\columnwidth]{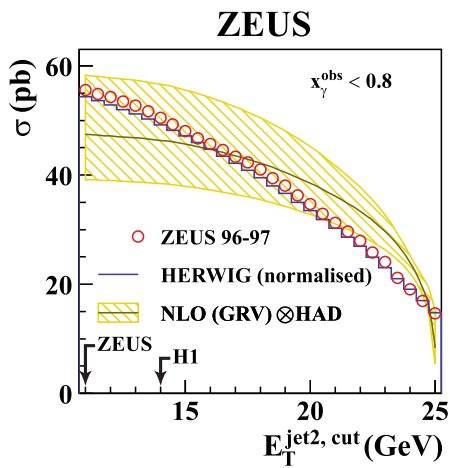}
  \end{center}
  \vspace{-0.8cm}
  \caption{
    ZEUS measurement of the dijet photoproduction cross section as a function
    of $E_{T}^{\rm Jet2, cut}$ for $25 < E_{T,1} < 35$~GeV and for 
    $x_\gamma^{\rm Obs} < 0.8$ (open circles).
    The HERWIG prediction is shown as the solid histogram and the result
    of the NLO QCD calculation is shown as the solid line.
    The theoretical uncertainty is shown as the hatched band.
    The positions of the cuts made by the H1 and the ZEUS collaborations
    are indicated.
  }
  \label{Fig:DijetXSect}
\end{figure}

The apparent discrepancy in the extent to which NLO QCD describes the
data is resolved by considering the dependence of the cross section on
the cut on $E_{T,{2}}$.
Consider the H1 measurement in the bin $25 < E_{T,{\rm max}} < 35$~GeV
and the ZEUS measurement in the bin $25 < E_{T,{1}}<35$~GeV.
The cross section for $E_{T,{1}}>25$~GeV is plotted as a
function of the cut on $E_{T,{2}}$ 
($E_{T}^{\rm jet2,cut}$) in figure \ref{Fig:DijetXSect}.
Also shown are the results obtained using the leading-log shower Monte
Carlo HERWIG and a NLO QCD calculation \cite{Ref:HERWIG,Ref:NLOQCDCalc}.
The HERWIG Monte Carlo, which has been normalised to the data, gives a
good description of the shape of the cross section.
The NLO QCD calculation, on the other hand, gives a good description
of the size of the cross section but is unable to describe the shape.
Note that the theoretical uncertainty is large and increases as
$E_{T}^{\rm jet2,cut}$ falls.
The difference in the level of agreement of the ZEUS and H1 data with
the calculation can now be explained by noting the position of the
cut applied on $E_{T,{2}}$. 
As shown in the figure, the H1 collaboration have chosen a value of
$E_{T}^{\rm jet2,cut}$ in a region in which NLO QCD is close to
the data, while the ZEUS collaboration cuts in a region where the NLO
QCD prediction lies significantly above the data.

The large data sets soon to be collected at HERA II will make a
substantial reduction in the experimental error possible if the
experimental collaborations can reduce the systematic uncertainty.
To extract quantitative information from this beautiful data will
require progress in reducing the theoretical uncertainty, perhaps
through going to a next-to-NLO calculation.

\section{Conclusions}
\label{Sect:Conclusions}

The papers submitted to the QCD sessions at this conference amply
demonstrate the breadth and depth of activity in the field.
At the end of the LEP era we have a detailed understanding of the
process $e^+e^- \rightarrow {\rm hadrons}$.
This understanding has been exploited to yield measurements of the
fundamental parameters of QCD, $\alpha_{\rm S}$ and the colour
factors at a precision of $\sim 1-5\%$ and $\sim 20\%$ respectively.
Lepton-nucleon deep inelastic scattering has provided measurements of
the partonic structure of the proton with a precision of $\sim 5\%$
for quarks and $\sim 15-20\%$ for the gluons.
Yet much remains to be accomplished if we are to achieve a complete,
quantitative, understanding of QCD.
The detailed measurements of diffraction in deep inelastic scattering
present a clear challenge and highlight the need to continue the
experimental and theoretical investigation of the transition from the
perturbative to the non-perturbative regime.

The coming years will see HERA and the Tevatron deliver large data
sets.
The experimental challenge will be to ensure that the systematic
uncertainties are reduced to match the statistical precision of the
data. 
The LEP measurements of $\alpha_{\rm S}$ demonstrate the precision
which can be achieved if theoretical uncertainties can be made at
least as small as those of the experiment.
At present the theoretical uncertainty on the majority of
hadron-induced cross sections is large compared to the anticipated
statistical precision. 
The ambition to develop a full, quantitative, understanding of QCD now
requires a broad and sustained programme of measurement and
interpretation. 
By energetically developing the already strong partnership between
theorists, phenomenologists and experimentalists we can work
confidently to achieve this ambition.

\section*{Acknowledgements}

I would like to thank the organisers for giving me the opportunity
to present this review.
I gratefully acknowledge the help, advice and support of my many 
colleagues, both experimentalists and theorists, who have freely 
discussed their results with me.
Finally, I would like to thank O.~Gonzalez, K.~Nagano and A.~Tapper
for help in the preparation of the figures and E.~Laenen for technical
support at the conference and patience through-out.


\begin{thebibliography}{99}
%
  \bibitem{Ref:ALEPH4Jet}
    ALEPH Collaboration, A.~Heister et al., CERN-EP/2002-029.
    Submitted to Eur. Phys. J. C.
%
  \bibitem{Ref:OPAL4Jet}
    OPAL Collaboration, G.~Abbiendi et al., Eur. Phys. J. C 20 (2001) 601.
%
  \bibitem{Ref:DELPHI4Jet}
    DELPHI Collaboration, P.~Abreu et al., Phys. Lett. B414 (1997) 401.
%
  \bibitem{Ref:DELPHIQGNch}
    DELPHI Collaboration, P.~Abreu et al., Phys. Lett. B 449 (1999) 383.
%
  \bibitem{Ref:DELPHIRGI}
    DELPHI Collaboration, O.~Pason et al., Abstract 228, paper
    contributed to this conference.  \\
    K.~Hamacher these proceedings.
%
  \bibitem{Ref:JADEPCFit}
    JADE Collaboration, Eur. Phys. J. C (2001) 199.
%
  \bibitem{REF:LEPEvtShpAlfs}
    G.~Dissertori, these proceedings and references therein.
%
  \bibitem{Ref:JADERunAlf}
    P.A.~Movilla~Fernandez, these proceedings.
%
  \bibitem{Ref:ZEUSInclJet}
    ZEUS Collaboration, Abstract 855,  paper contributed to this
    conference. \\
    M.~Sutton, these proceedings.
%
  \bibitem{Ref:CDFInclJet}
    CDF Collaboration, T.~Affolder et al., Phys. Rev. Lett. 88 (2002) 042001.
%
  \bibitem{Ref:CDFJetPap}
    CDF Collaboration, T.~Affolder et al., Phys. Rev. D 64 (2001) 032001.
%
  \bibitem{Ref:WorldAverage}
    PDG, Phys. Rev. D 66 (2002) 010001;                               \\
    S.~Bethke, J. Phys. G 26 (2000) R27.
%
  \bibitem{Ref:H1Pos1}
    H1 Collaboration, C.~Adloff et al., Eur. Phys. J. C 21 (2001) 33.
%
  \bibitem{Ref:H1Pos2}
    H1 Collaboration, Abstract 978, paper contributed to this
    conference. \\
    Z. Zhang, these proceedings.
%
  \bibitem{Ref:ZF2em}
    ZEUS Collaboration, S.~Chekanov et al., Eur. Phys. C 21 (2001)
    443.
%
  \bibitem{Ref:FixedTarget}
    NMC Collaboration, M.~Arneodo et al., Nucl. Phys. B 483 (1997) 3;   \\
    BCDMS Collaboration, A.C.~Benvenuti et al., Phys. Lett. B 223
    (1989) 485;                                                         \\
    E665 Collaboration, M.R. Adams et al., Phys. Rev. D 54 (1996) 3006.
%
  \bibitem{Ref:DGLAP}
    Y.L.~Dokshitzer, Sov. Phys. JETP 46 (1977) 641;                     \\
    V.N.~Gribov and L.N.~Lipatov, Sov. J. Nucl. Phys. 15 (1972) 438
    and 675;                                                            \\
    G.~Alterelli and G.~Parisi, Nucl. Phys. B 126 (1977) 298.
%
  \bibitem{Ref:CTEQ6}
    J.~Pumplin et al., JHEP 07 (2002) 012.
%
  \bibitem{Ref:MRST2001}
    A.D.~Martin et al., Eur. Phys. J. C 23 (2002) 73.    
%
  \bibitem{Ref:H1NLOQCDFits}
    H1 Collaboration, C.~Adloff et al., Eur. Phys. J. C 21 (2001) 33;   \\
    H1 Collaboration, Abstract 978, paper contributed to this
    conference. \\
    Z.~Zhang, these proceedings.
%
  \bibitem{Ref:ZNLOQCDFit}
    ZEUS Collaboration, C.~Chekanov et al., DESY-02-105, submitted to
    Phys. Rev. D.
%
  \bibitem{Ref:H1Lambda}
    H1 Collaboration, C.~Adloff et al., Phys. Lett. B 520 (2001) 183.
%
  \bibitem{Ref:ZEUSLambda}
    J.~Breitweg et al., Eur. Phys. J. C 7 (1999) 609.
%
  \bibitem{Ref:H1Ele}
    H1 Collaboration, C.~Adloff et al., Eur. Phys. J. C 19 (2001) 269.
%
  \bibitem{Ref:ZPos1NC}
    ZEUS Collaboration, J.~Breitweg et al., Eur. Phys. J. C 11 (1999) 427.
%
  \bibitem{Ref:ZPos1CC}
    ZEUS Collaboration, J.~Breitweg et al., Eur. Phys. J. C 12 (2000) 411.
%
  \bibitem{Ref:ZEleNC}
    ZEUS Collaboration, S.~Chekanov et al., DESY-02-113, submitted to 
    Eur. Phys. J. C.
%
  \bibitem{Ref:ZEleCC}
    ZEUS Collaboration, S.~Chekanov et al., Phys. Lett. B 539 (2002) 197.
%
  \bibitem{Ref:ZPos2}
    E.~Rizvi in Proceedings of the EPS International Conference on
    High Energy Physics, Budapest 2001 (D.~Howvarth, P.~Levai,
    A.~Patkos, eds.), JHEP (http://jhep.sissa.it/) Proceedings
    Section, PrHEP-hep2001/011 and references therein.
%
  \bibitem{Ref:D0InclJets}
    D0 Collaboration, B.~Abbott et al., Phys. Rev. Lett. 86 (2001)
    1707, Phys. Rev. D 64 (2001) 032003.
%
  \bibitem{Ref:CTEQ5}
    CTEQ Collaboration, H.L~Lai et al., Eur. Phys. J. C 12 (2000) 375.
%
  \bibitem{Ref:MRSTNNLO}
    A.D.~Martin et al., Phys. Lett. B 531 (2002) 216.
%
  \bibitem{Ref:LEPF2Gam}
    R.~Nisius, these proceedings and references therein.
%
  \bibitem{Ref:F2GamAlf}
    S.~Albino, M.~Klasen, Phys. Rev. Lett. 89 (2002) 122004.
%
  \bibitem{Ref:ZEUSdsdt}
    ZEUS Collaboration, Abstract 566,  paper contributed to this
    conference. 
%
  \bibitem{Ref:Collins}
    J.C.~Collins, Phys. Rev. D 57 (1998) 3051 and erratum-ibid. D 61
    (2000) 019902.
%
  \bibitem{Ref:H1PomFit}
    H1 Collaboration, Abstract 980,  paper contributed to this
    conference.                  \\
    F.P.~Schilling, these proceedings.
%
  \bibitem{Ref:CDM}
    J.~Bartels et al., Eur. Phys. J. C 7 (1999) 443.
%
  \bibitem{ZEUS:dsdMX}
    ZEUS Collaboration, S.~Chekanov et al., Eur. Phys. J. C 
    DOI 10.1140/s10052-002-1003-1.                              \\
    ZEUS Collaboration, Abstract 822,  paper contributed to this
    conference.
%
  \bibitem{Ref:H1PhotDiJets}
    H1 Collaboration, C.~Adloff et al., Eur. Phys. J. C 25 (2002) 1.
%
  \bibitem{Ref:ZPhotDiJets}
    ZEUS Collaboration, S.~Chekanov et al., Eur. Phys. J. C 23 (2002) 4. 
%
  \bibitem{Ref:HERWIG}
    G.~Marchesini et al., Comp. Phys. Comm. 67 (1992) 465.
%
  \bibitem{Ref:NLOQCDCalc}
    S.~Frixione, Z.~Kunszt and A.~Signer, Nucl. Phys. B 467 (1996) 399; \\
    S.~Frixione, Nucl. Phys. B 507 (1997) 295;                          \\
    S.~Frixione, G.~Ridolfi, Nucl. Phys. B 507 (1997) 315.
%
\end{thebibliography}
\end{document}